\documentclass[prd,aps,showpacs,preprintnumbers,amsmath,amssymb,11pt]{revtex4}
\usepackage{fleqn,epsf}
\usepackage{graphicx}
\usepackage{psfrag}
\newcommand{\be}{\begin{equation}}
\newcommand{\ee}{\end{equation}}
\newcommand{\bea}{\begin{eqnarray}}
\newcommand{\eea}{\end{eqnarray}}

\def\lsim{\raise0.3ex\hbox{$\;<$\kern-0.75em\raise-1.1ex\hbox{$\sim\;$}}}
\def\gsim{\raise0.3ex\hbox{$\;>$\kern-0.75em\raise-1.1ex\hbox{$\sim\;$}}}
\begin{document}
%\begin{titlepage}
%\baselineskip=0.33in
%\thispagestyle{empty}
% declarations for front matter

\title{Two-Photon and Two-gluon Decays of \\$0^{++}$ and $2^{++}$ 
  P-wave Heavy Quarkonium States\footnote{
Talk given at the 
{\em QCD@Work 2010 International Workshop on QCD: Theory and
  Experiment}, Martina Franca, Valle d'Istria, Italy, 20--23 June 2010}}
\author{
T. N. Pham}
\affiliation{
 Centre de Physique Th\'{e}orique, CNRS \\ 
Ecole Polytechnique, 91128 Palaiseau, Cedex, France }
\date{\today}

\begin{abstract}
By neglecting the relative quark momenta  in the propagator term,
the two-photon and two-gluon
decay amplitude of heavy quarkonia states
can be written as  a local
heavy quark field operator matrix element which could be obtained from 
other processes or computed with  QCD sum rules technique or lattice 
simulation, as shown in a recent  work on $\eta_{c,b}$ two-photon
decays. In this talk, I
would like to discuss a similar calculation on  $P$-wave
$\chi_{c0,2}$ and $\chi_{b0,2}$ two-photon decays.
We  show that the effective Lagrangian for the two-photon decays of 
the $P$-wave $\chi_{c0,2}$ and $\chi_{b0,2}$ is 
given by  the heavy quark energy-momentum tensor local operator and its
trace, the  $\bar{Q}Q$ scalar density. A simple expression
for $\chi_{c0}$  two-photon and two-gluon decay rate in terms of  
the $f_{\chi_{c0}}$ decay constant, similar to that
of $\eta_{c}$ is obtained. From the existing
 QCD sum rules value for $f_{\chi_{c0}}$, we get $5\rm\,keV$ for 
the $\chi_{c0}$ two-photon width, somewhat larger than measurement.  
\end{abstract}


\maketitle

\section{Introduction}
First of all, I would like to dedicate this talk to the memory of
Professor Giuseppe Nardulli, who, with great kindness and generosity 
has initiated the long and fruitful collaboration I have with 
the members of the Physics Department and INFN at the University of
Bari.

In the non-relativisitic bound state calculation
\cite{Barbieri,Brambilla}, the two-photon and
two-gluon decay rates 
for  $P$-wave quarkonium states depend on the derivative
of the spatial wave function at the origin which has to be extracted from
potential models, unlike the two-photon decay rate
 of $S$-wave  $\eta_{c}$ and $\eta_{b}$ quarkonia which can be 
predicted from the corresponding $J/\psi$ and $\Upsilon$ leptonic widths 
using heavy quark spin symmetry(HQSS) \cite{Lansberg}, there is 
no similar  prediction for the $P$-wave $\chi_{c}$ and $\chi_{b}$
states and all the existing theoretical values for the decay rates are based
on potential model calculations \cite{Barbieri,Godfrey,Barnes,Bodwin,Gupta,Munz,Huang,Ebert,Schuler,Crater,LWang,Laverty}.

Since the matrix element of a heavy quark local operator between 
the vacumm and $P$-wave quarkonium state is also given, in bound 
state description, by the derivative of the spatial wave function 
at the origin, one could express the $P$-wave quarkonium two-photon 
and two-gluon decay amplitudes in terms of the matrix element of 
a local operator with the appropriate quantum number, like the 
heavy quark $\bar{Q}Q$ scalar density
or axial vector current $\bar{Q}\gamma_{\mu}\,\gamma_{5}Q$. We have thus  
an effective Lagrangian for  the two-photon and two-gluon decays 
of $P$-wave quarkonia in terms of heavy quark field operator 
instead of the traditional bound state description in terms of the
wave function. 
This effective Lagrangian can be derived 
in a simple manner by neglecting the relative quark momentum 
in the heavy quark propagator  as in non-relativistic bound
state calculation. In this talk, I would like to report on  a recent 
work \cite{Lansberg2} using the effective Lagrangian approach
to describe the two-photon and two gluon decays of $P$-wave 
heavy quarkonia state, similar to that for $S$-wave 
quarkonia \cite{Lansberg} . This was stimulated by the recent new 
CLEO measurements \cite{CLEO,PDG} 
of the two-photon decay rates of the charmonium $P$-wave $0^{++}$,
 $\chi_{c0}$ and  $2^{++}$ $\chi_{c2}$ states. We obtain 
 an effective Lagrangian for $P$-wave quarkonium decays in terms of 
the heavy quark energy-momentum tensor and  its trace and that 
the two-photon and two-gluon decay rates of $\chi_{c0,2}$ and 
$\chi_{b0,2}$  can be expressed in terms of the decay constants 
$f_{\chi_{c0}}$ and $f_{\chi_{b0}}$, similar to that for  $\eta_{c}$ and
$\eta_{b}$,  which are given, respectively, by 
$f_{\eta_{c}}$ and $f_{\eta_{b}}$. Then a calculation of 
$f_{\chi_{c0}}$ and $f_{\chi_{b0}}$ by  sum rules 
technique \cite{Novikov,Reinders} or lattice
simulation \cite{Dudek,Chiu}  would give us a prediction of 
the $P$-wave quarkonia decay rates. In fact, as shown below,
 $f_{\chi_{c0}}$ obtained in \cite{Novikov} implies a value of
 $5\rm\,keV$ for the $\chi_{c0}$ two-photon width, somewhat larger 
than measurement. In the following I will present only the 
main results, as more details 
are given in the published paper \cite{Lansberg2}. 
\begin{figure}[h]
\centering
\includegraphics[height=2.5cm,angle=0]{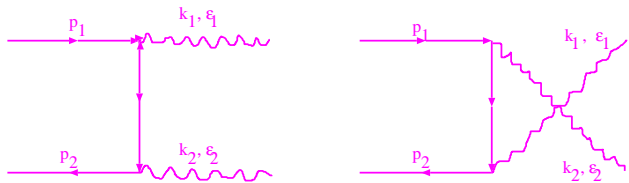}
\caption{Diagrams for $Q\bar{Q} $ annihilation to two photons.}
\label{fig1}
\end{figure}

\section{Effective Lagrangian for $\chi_{c0,2}\to \gamma\gamma$ and $\chi_{b0,2}\to \gamma\gamma$}

By neglecting term containing the
relative quark momenta $q$ in the quark propagator\cite{Kuhn} 
 ($Q^{2}_{c,b} $ being the heavy quark charge), the  
 $P$-wave part of the $c\bar{c} \to
\gamma\gamma, gg $ and $b\bar{b} \to \gamma\gamma, gg $ amplitudes represented
by  diagrams in Fig. \ref{fig1} are
\be
{\cal M}(Q\bar{Q}\to \gamma\gamma)=-e^{2}Q^{2}_{c,b}\frac{A_{\mu\nu}\bar{v}(p_{2})T_{\mu\nu}u(p_{1})}
{[(k_{1}-k_{2})^{2}/4 -m_{Q}^{2}]^{2}}
\label{Amn}
\ee
with $A_{\mu\nu}$ the photon part of the amplitude  and the
heavy quark part $T_{\mu\nu}$ given by
\bea
A_{\mu\nu}&=& -2\epsilon_1\cdot k_{2}\epsilon_{2\mu}k_{1\nu}+2\epsilon_1\cdot\epsilon_2 k_{2\mu}k_{1\nu}\nonumber \\
&&-2\epsilon_2\cdot k_{1}\epsilon_{1\mu}k_{2\nu}+(k_{1}\cdot k_{2})(\epsilon_{1\mu}\epsilon_{2\nu}+\epsilon_{2\mu}\epsilon_{1\nu}) \\
T_{\mu\nu}&=& (q_{1\mu}-q_{2\mu})\gamma_{\nu}
\label{T}
\eea
which can  be obtained directly from the following  effective Lagrangian 
for two-photon and two-gluon
decay of $P$-wave heavy quarkonia  states
\bea
\kern-0.5cm {\cal L}_{\rm eff}(Q\bar{Q}\to \gamma\gamma)&=&-ic_{1}A_{\mu\nu}
\bar{Q}(\overrightarrow\partial_{\mu}-\overleftarrow\partial_{\mu})\gamma_{\nu}Q\\
\label{Leffc}
\kern-0.5cm c_{1}&=& -e^{2}Q^{2}_{c,b}[(k_{1}-k_{2})^{2}/4 -m_{Q}^{2}]^{-2}\nonumber
\eea
With   the matrix element of
$\theta_{Q\mu\nu}=\bar{Q}(\overrightarrow\partial_{\mu}-\overleftarrow\partial_{\mu})\gamma_{\nu}Q$ between the vacuum and $\chi_{c0,2}$ or $\chi_{b0,2}$
given by ($Q^{2}=M^{2}$)
\bea
<0|\theta_{Q\mu\nu}|\chi_{0}> &=& T_{0}M^{2}(-g_{\mu\nu} + Q_{\mu}Q_{\nu}/M^{2}),\nonumber\\
<0|\theta_{Q\mu\nu}|\chi_{2}> &=& -T_{2}M^{2}\epsilon_{\mu\nu}.
\label{Tmn}
\eea
  The two-photon decay amplitudes are then easily obtained~:
\bea
{\cal M}(\chi_{0}\to \gamma\gamma)&=& -e^{2}Q^{2}_{c,b}\frac{T_{0}A_{0}}{[M^{2}/4 +m_{Q}^{2}]^{2}}\\
\label{chi0}
{\cal M}(\chi_{2}\to \gamma\gamma)&=& -e^{2}Q^{2}_{c,b}\frac{T_{2}A_{2}}{[M^{2}/4 +m_{Q}^{2}]^{2}}
\label{chi2}
\eea
with $T_{2}=\sqrt{3}T_{0}$ from HQSS and
\bea
A_{0} &=&({3\over 2})M^{2}(M^{2}\epsilon_{1}\cdot\epsilon_{2}-2\epsilon_{1}\cdot k_{2}\epsilon_{2}\cdot k_{1})\\
\label{A0}
A_{2}&=&M^{2}\epsilon_{\mu\nu}[M^{2}\epsilon_{1\mu}\epsilon_{2\nu}
-2(\epsilon_{1}\cdot k_{2}\epsilon_{2\mu}k_{1\nu} + \epsilon_{2}\cdot
k_{1}\epsilon_{1\mu}k_{2\nu}\nonumber \\
&& + \epsilon_{1}\cdot \epsilon_{2}k_{1\mu}k_{2\nu})]
\label{A2}
\eea

  For QCD sum rules calculation
or lattice simulation, it is simpler to compute the trace of the 
energy-momentum tensor $\theta_{Q\mu\mu}$ given 
by $2m_{Q}\bar{Q}Q $. We have then
\be
\bar{v}(p_{2})T_{\mu\mu}u(p_{1}) = 2\,m_{Q}\,\bar{v}(p_{2})u(p_{1})
\label{Tmm}
\ee
  The problem of computing the two-photon or 
two-gluon decay amplitude of $\chi_{c0,2}$ and $\chi_{b0,2}$ states 
is reduced to computing the decays constants $f_{\chi_{c0}}$ and
 $f_{\chi_{b0}}$ defined as 
\be
<0|\bar{Q}Q|\chi_{0}> = m_{\chi_{0}}f_{\chi_{0}}
\label{fchi}
\ee
 Thus  $ T_{0} $ is given directly in terms of $f_{\chi_{0}} $
without using the derivative of the $P$-wave
spatial wave function at the origin.
\be
 T_{0}= {f_{\chi_{0}}\over 3} 
\label{C0}
\ee

  Thus by comparing the expression for 
$\chi_{c0}$ and $\eta_{c}$ we could already have some estimate for the
$\chi_{c0}$ two-photon and two-gluon decay rates. For $f_{\chi_{c0}}$
of $O(f_{\eta_{c}})$, one would expect
 $\Gamma_{\gamma\gamma}(\chi_{c0})$ to be in the range of a few keV. 

  The decay rates of $\chi_{c0,2}$, $\chi_{b0,2}$
states can now be obtained in terms of the decay constant
$f_{\chi_{0}}$. We have~:
\bea
&&\kern-0.9cm \Gamma_{\gamma\gamma}(\chi_{c0})= \frac{4 \pi Q_c^4
   \alpha^2_{em}M_{\chi_{c0}}^{3} f_{\chi_{c0}}^2}{(M_{\chi_{c0}}+ b)^{4}}\left[1 + B_{0}(\alpha_{s}/\pi)\right],
\label{R0b}\\
&&\kern-0.9cm \Gamma_{\gamma\gamma}(\chi_{c2})= \left({4 \over 15}\!\right)\frac{4 \pi Q_c^4 \alpha^2_{em}M_{\chi_{c2}}^{3} f_{\chi_{c0}}^2}{(M_{\chi_{c2}}+ b)^{4}}\left[1 + B_{2}(\alpha_{s}/\pi)\right]
\label{R2}
\eea
where $B_{0}= \pi^{2}/3 -28/9$ and $B_{2}=  -16/3$ are NLO
QCD radiative corrections \cite{Barbieri2,Kwong,Mangano}

  This expression is  similar to that for $\eta_{c}$~:
\be
 \Gamma_{\gamma\gamma}(\eta_{c})= \frac{4 \pi Q_c^4
   \alpha^2_{em}M_{\eta_{c}} f_{\eta_{c}}^2}{(M_{\eta_{c}}+
   b)^{2}}\left[1 -{\alpha_{s}\over \pi}{(20-\pi^{2})\over 3}\right]
\label{Retac}
\ee
  The two-gluon decay rates are~:
\bea
&&\kern-0.9cm \Gamma_{gg}(\chi_{c0})= \left({2\over 9}\right)\frac{4 \pi 
   \alpha^2_{s}M_{\chi_{c0}}^{3} f_{\chi_{c0}}^2}{(M_{\chi_{c0}}+ b)^{4}}[1 + C_{0}(\alpha_{s}/\pi)],
\label{R0gg}\\
&&\kern-0.9cm \Gamma_{gg}(\chi_{c2})= \left({4 \over 15}\right)\!\left({2\over 9}\right)\!\frac{4 \pi  \alpha^2_{s}M_{\chi_{c2}}^{3} f_{\chi_{0}}^2}{(M_{\chi_{c2}}+ b)^{4}}[1 + C_{2}(\alpha_{s}/\pi)]
\label{R2gg}
\eea
where $C_{0}= 8.77$ and $C_{2}= -4.827$ are NLO QCD radiative corrections.
As with the two-photon decay rates,  the expressions for two-gluon decay 
rates are similar to that for $\eta_{c}$:
\be
 \Gamma_{gg}(\eta_{c})=\left({2\over 9}\right) \frac{4 \pi
   \alpha^2_{s}M_{\eta_{c}} f_{\eta_{c}}^2}{(M_{\eta_{c}}+
   b)^{2}}\left[1 +4.8{\alpha_{s}\over \pi}\right]
\label{Retacgg}
\ee
  In a bound state calculation, using the relativistic spin projection 
operator \cite{Kuhn,Guberina} , $f_{\eta_{c}} $ and $f_{\chi_{0}} $ 
are given by
\bea
&&f_{\eta_{c}} = \sqrt{\frac{3}{32\,\pi\,m_{Q}^{3}}}\,{\cal R}_{0}(0)\, (4\,m_{Q}) \ , \qquad
\label{fetac}\\
&&f_{\chi_{0}}= 12\sqrt{{3\over (8\pi m_{Q})}}\left({{\cal R}_{1}^{\prime}(0)\over M}\right)\\
\label{fchic}
\eea
which gives the decay amplitudes in agreement with the original
calculation \cite{Barbieri}. 

Comparing with $f_{\eta_{c}} $, we have
\be
f_{\chi_{c0}} = 6\left({{\cal R}^{\prime}_{1}(0)\over {\cal
 R}_{0}(0)M}\right)f_{\eta_{c}}.
\label{fSP}
\ee
which  becomes comparable to $f_{\eta_{c}}$.

 Thus by comparing the expression for 
$\chi_{c0}$ and $\eta_{c}$ we could already have some estimate for the
$\chi_{c0}$ two-photon and two-gluon decay rates. For $f_{\chi_{c0}}$
of $O(f_{\eta_{c}})$, one would expect
 $\Gamma_{\gamma\gamma}(\chi_{c0})$ to be in the range of a few keV. 
  As shown in Table 1, the predicted two-photon width
of $\chi_{c0}$ from the sum rules value  $f_{\chi_{c0}}=357\,\rm MeV$
\cite{Novikov} is however almost twice  the CLEO value, but
possibly with large theoretical uncertainties in sum rules calculation 
 as to be expected, while a recent calculation \cite{Colangelo} implies 
a larger decay rates for $\chi_{c0}$.
  The measured  ratio
$\Gamma_{\gamma\gamma}(\chi_{c2})/\Gamma_{\gamma\gamma}(\chi_{c0})$ 
is then $\approx 0.24\pm 0.09$, somewhat bigger than the predicted 
value of about $0.14$ as shown in Table 1 together with the 
CLEO measurement of the decay rates \cite{CLEO} which gives $(2.53\pm 0.37\pm
0.26)\rm\,keV$ and $(0.60\pm 0.06\pm 0.06)\rm\,keV$ for 
$\chi_{c0}$ and  $\chi_{c2}$ respectively.
\begin{table}[h]
\begin{tabular}{|c|c|c|c|}
\hline
\hline
 Reference
 &$\Gamma_{\gamma\gamma}(\chi_{c0})$(keV)&$\Gamma_{\gamma\gamma}(\chi_{c2})$(keV)&$R={\Gamma_{\gamma\gamma}(\chi_{c2})\over
 \Gamma_{\gamma\gamma}(\chi_{c0})}$ \\ 
\hline
Barbieri\cite{Barbieri}&$3.5$ &$0.93$ &$ 0.27$\\
Godfrey\cite{Godfrey}&$1.29$  &$0.46$ & $0.36$\\
Barnes\cite{Barnes}&$1.56$ &$0.56$&$0.36$ \\
Gupta\cite{Gupta} &$6.38$&$0.57 $ &$0.09 $\\
M\"unz\cite{Munz} &$1.39\pm 0.16 $ &$0.44\pm 0.14 $&$0.32^{+0.16}_{-0.12}$\\
Huang\cite{Huang} &$3.72\pm 1.10 $ &$0.49\pm 0.16 $&$0.13^{+0.11}_{-0.06}$\\
Ebert\cite{Ebert} &$2.90 $ &$0.50 $&$0.17$\\
Schuler\cite{Schuler} &$2.50 $ &$0.28 $&$0.11$\\
Crater\cite{Crater} &$3.34-3.96 $ &$0.43-0.74 $&$0.13-0.19$\\
Wang\cite{LWang} &$3.78 $ &$- $&$-$\\
Laverty\cite{Laverty} &$1.99-2.10 $ &$0.30-0.73 $&$0.14-0.37$\\
This work  &$5.00 $ &$0.70 $&$0.14$\\
Exp(CLEO)\cite{CLEO} &$2.53\pm 0.37\pm 0.26 $ &$0.60\pm 0.06\pm 0.06 $&$0.24\pm 0.04\pm 0.03$\\
Exp(Average)\cite{CLEO} &$2.31\pm 0.10\pm 0.12 $ &$0.51\pm 0.02\pm 0.02 $&$0.20\pm 0.01\pm 0.02$\\
\hline
\end{tabular}
\caption{Potential model predictions for $\chi_{c0,2}$
two-photon widths compared with this work.}
\end{table}

The two-photon  $\chi_{c0,2},\chi_{c0,2}^{\prime}$ branching ratios are independent of 
$f_{\chi_{c0}}$
\bea
&& {\cal B}(\chi_{c0},\chi_{c^{\prime}0}\to \gamma\gamma) = \frac{9}{2}\,Q_c^4\,\frac{\alpha^{2}_{em}}{\alpha^{2}_{s}}\left(1+(B_{0}-C_{0})\frac{\alpha_s}{\pi}\right)\label{BR2gchi0}\\
&&{\cal B}(\chi_{c2},\chi_{c^{\prime}2}\to \gamma\gamma) = \frac{6}{5}\,Q_c^4\,\frac{\alpha^{2}_{em}}{\alpha^{2}_{s}}\left(1+(B_{2}-C_{2})\frac{\alpha_s}{\pi}\right)
\label{BR2gchi2}
\eea
with $B_{0}= \pi^{2}/3 -28/9, B_{2}=  -16/3, C_{0}= 8.77, C_{2}= -4.827$.
Apart from QCD radiative correction factors, the expressions for 
branching ratios are
very similar to that for $\eta_{c}$ and  $\eta_{c}^{\prime}$:
\be
 {\cal B}(\eta_{c},\eta_{c^{\prime}}\to \gamma\gamma) = \frac{9}{2}\,Q_c^4\,\frac{\alpha^{2}_{em}}{\alpha^{2}_{s}}\left(1-8.2\,\frac{\alpha_s}{\pi}\right)
\label{BR2get}
\ee
with $\alpha_{s} $ evaluated at the appropriate scale.

  For  $\alpha_{s}=0.26 $, 
  $ {\cal B}(\eta_{c}\to  \gamma\gamma)= 3.6\times 10^{-4} $ to be 
compared with the measured
value of $(2.8\pm 0.9)\times 10^{-4} $ \cite{PDG}, but this prediction is
rather sensitive to $\alpha_{s} $, for example, with $\alpha_{s}=0.28 $, 
one would get $ {\cal B}(\eta_{c}\to \gamma\gamma)= 2.95\times 10^{-4}$, 
in better agreement with the measured value of
 $(2.4^{+1.1}_{-0.9})\times 10^{-4} $ and for $\chi_{c0,2}$, the
predicted two-photon branching ratios would be 
 $3.45\times 10^{-4}$ and $4.45 \times 10^{-4} $ 
compared with the measured values of  $(2.35 \pm 0.23)\times 10^{-4} $
and $(2.43 \pm 0.18)\times 10^{-4} $, for $\chi_{c0}$ and $\chi_{c2}$
respcetively. The predicted branching ratio for $\chi_{c2}$
is rather large and one would need $\alpha_{s} =0.36$ to bring the predicted
value closer to experiment.

  Recently,  the $Z(3930)$ state above $D\bar{D}$ threshold found 
by  Belle \cite{Belle2} with  mass $(3928\pm 5\pm 2)\rm\,MeV$  
and  width  $(29\pm 10({\rm stat})\pm 2({\rm sys}))\rm\,MeV$,
consistent with  $\chi^{\prime}_{c2}$, seems to be confirmed by 
the observation of a similar state  by  BaBar \cite{Babar2}, with 
mass $(3926.7\pm 2.7\pm 1.1)\rm\,MeV$ and width 
$(21.3\pm 6.8\pm 3.6)\rm\,MeV$. Belle \cite{Belle2} gives
$\Gamma_{\gamma\gamma}(\chi^{\prime}_{c2})\times {\cal
  B}(D\bar{D})=(0.18\pm 0.05\pm 0.03)\rm\,keV$ while   
 Babar \cite{Babar2} gives
$\Gamma_{\gamma\gamma}(\chi^{\prime}_{c2})\times {\cal
  B}(D\bar{D})=(0.24\pm 0.05\pm 0.04)\rm\,keV$ for this state.
If taken to be the $2P$ excited state
 $\chi^{\prime}_{c2}$ and assuming   $ {\cal B}(D\bar{D})\approx 0.70-1$ 
\cite{Colangelo2,Swanson,Chao}, one would get
 $\Gamma_{\gamma\gamma}(\chi^{\prime}_{c2})=(0.18-0.24\pm
0.05\pm 0.03)\rm\,keV$ . This implies  
  $f_{\chi^{\prime}_{c0}}\simeq (195-225)\rm\,MeV$ and 
 $\Gamma_{gg}(\chi^{\prime}_{c0})$ in the range $(5-10) \rm\,MeV$.

  For $\chi_{b 0,2}$ potential model calculations similar to that for
$\chi_{c0,2}$, gives the two-photon width about $1/10$ of 
that for $\eta_{b}$ , which implies $f_{\chi_{b 0}}= f_{\eta_{b}}/3$, 
smaller than Cornell potential \cite{Quigg} value 
$f_{\chi_{b 0}}= 0.46\,f_{\eta_{b}}$.
%\newpage 
\section{Remark on the $\eta_{c}^{\prime}$ two-photon decays}
Since the predicted two-photon branching ratios for 
$\chi_{c0,2},\chi_{c0,2}^{\prime}$ and for $\eta_{c}, \eta_{c}^{\prime}$
are similar and independent of the decay constants, apart from QCD 
radiative corrections, as seen in Eq. (\ref{BR2gchi0}-\ref{BR2gchi2}) 
and Eq. (\ref{BR2get}), one expects a large two-photon decay rates 
for $\eta_{c}^{\prime}$, it would be  relevant here to mention the 
problem of the $\eta_{c}^{\prime}\to \gamma\gamma$ decay rate 
\cite{Lansberg,Pham}. The small value of 
$\Gamma_{\gamma \gamma}(\eta_{c}') = (1.3 \pm 0.6) \,{\rm keV} $ given
previously by CLEO \cite{Asner}  is obtained by assuming 
${\cal B}(\eta_{c}' \to K_{S}K\pi) \approx {\cal B}(\eta_{c} \to
K_{S}K\pi) $. However, with the recent BaBar  measurement of the 
ratio \cite{Babar}
\bea
R(\eta_{c}(2S)K^{+}/\eta_{c}K^{+})&=&\frac{{\cal B}(B^{+}\to \eta_{c}(2S)K^{+})\times {\cal B}(\eta_{c}(2S)\to
K\bar{K}\pi )}{{\cal B}(B^{+}\to \eta_{c}K^{+})\times {\cal B}(\eta_{c}\to
K\bar{K}\pi )}\nonumber \\
&=&0.096^{+0.020}_{-0.019}({\rm stat})\pm 0.025({\rm syst})
\label{Rbabar}
\eea
and the Belle measurement \cite{Belle}
\be
{\cal B}(B^{+}\to \eta_{c}K^{+})\times {\cal B}(\eta_{c}\to
K\bar{K}\pi )=(6.88\pm 0.77^{+0.55}_{-0.66})\times 10^{-5} 
\label{babar}
\ee
BABAR obtains \cite{Babar}~
\be
  {\cal B}(\eta_{c}^{\prime} \to K_{S}K\pi) = (1.9\pm 0.4(\rm stat)\pm
  1.1(\rm syst))\% .
\label{babar1}
\ee
as quoted  by CLEO \cite{CLEO2}. This  new BABAR value for
${\cal  B}(\eta_{c}^{\prime} \to K_{S}K\pi) $ is considerably  smaller 
than the corresponding value 
${\cal B}(\eta_{c} \to K_{S}K\pi) = (7.0\pm 1.2)\%$ \cite{PDG} for $\eta_{c}$.

  Thus with the BaBar result for ${\cal  B}(\eta_{c}^{\prime} \to K_{S}K\pi) $  and the  CLEO measurement \cite{Asner}
\be
R(\eta_{c}^{\prime}/\eta_{c}) = \frac{\Gamma_{\gamma \gamma}(\eta_{c}^{\prime})\times
  {\cal B}(\eta_{c}^{\prime} \to K_{S}K\pi)}{\Gamma_{\gamma \gamma}(\eta_{c})\times
  {\cal B}(\eta_{c} \to K_{S}K\pi)} = 0.18\pm0.05\pm 0.02
\label{CLEO}
\ee
one would get \cite{CLEO2}
\be
\Gamma(\eta_{c^{\prime}}\to \gamma\gamma)= (4.8\pm 3.7)\, \rm keV  
\ee
in agreement with the predicted value
\be
\Gamma(\eta_{c^{\prime}}\to \gamma\gamma)= (4.1\pm 2.3)\, \rm keV 
\label{eta2gg}
\ee
while the assumption of  near equality of the $K_{S}K\pi$ branching ratios
for $\eta_{c}$ and $\eta_{c}^{\prime}$~
\be
{\cal B}(\eta_{c}' \to K_{S}K\pi) \approx {\cal B}(\eta_{c} \to K_{S}K\pi)
\label{KKp}
\ee
and the  Belle  ratio \cite{Choi} 
\be
\kern-0.8cm R(\eta_{c}^{\prime}K/\eta_{c}K)=\frac{{\cal B}(B\to K\eta_{c}(2S)
\times {\cal B}(\eta_{c}(2S)\to
  K_{S}K^{-}\pi^{+})}{{\cal B}(B^{0}\to K\eta_{c})\times {\cal
    B}(\eta_{c}\to K_{S}K^{-}\pi^{+})} = 0.38\pm 0.12 \pm 0.05
\label{Belle}
\ee
would lead to \cite{CLEO2}
\be
\Gamma_{\gamma \gamma}(\eta_{c}^{\prime}) = (1.3 \pm 0.6) \,{\rm keV}
\label{etac'}
\ee
which is rather small compared with  the predicted value given 
in Eq. (\ref{eta2gg}) above.

\section{Conclusion}

  In conclusion, we have derived an  effective Lagrangian  
for  $\chi_{c0,2}$ and $\chi_{b0,2}$ two-photon and 
two-gluon  in terms of the decay constants $f_{\chi_{c,b0}}$, 
similar to that for $\eta_{c,b}$ in terms of $f_{\eta_{c,b}}$. 

  Existing sum rules
calculation,  however produces a 
two-photon width about $5\,\rm keV$, somewhat bigger than the 
CLEO measured value. It remains
to be seen whether a better determination of $f_{\chi_{c0}}$
from lattice simulation or QCD sum rules calculation 
could bring the $\chi_{c0,2}$ two-photon decay rates closer to experiments
or higher order QCD radiative corrections and large relativistic corrections
are needed to explain the data.

 The  problem of two-photon width of $\eta_{c}^{\prime}$ would go 
away if more data could confirm the small BaBar value for 
$ {\cal B}(\eta_{c}^{\prime} \to K_{S}K\pi) $ compared with
$ {\cal B}(\eta_{c} \to K_{S}K\pi)$. 

 As  relativistic corrections should be small for $P$-wave bottomia 
$\chi_{b0,2}$ states,  two-photon and two-gluon decays could provide 
a  test of QCD  and a determination of $\alpha_{s}$ at the 
the $m_{b}$ mass scale. 

\section{Acknowledgments}
I would like to thank  P. Colangelo, F. De Fazio and
the  organizers  for  generous support and warm hospitality 
extended to me at Martina Franca. This work was supported in part by the EU 
contract No. MRTN-CT-2006-035482, "FLAVIAnet".

\end{document}